# Critical Current Density and AC Magnetic Susceptibility of High-quality FeTe$_{0.5}$Se$_{0.5}$ Superconducting Tapes

Xin Zhou [1, #], Wenjie Li [1, 2, #], Qiang Hou[1], Wei Wei [1], Wenhui Liu [1], Ke Wang[1], Xiangzhuo Xing [3], Linfei Liu [4], Jun-Yi Ge[5], Yanpeng Qi[6], Huajun Liu[7], Li Ren[8], Tsuyoshi Tamegai[2], Yue Sun [1*] and Zhixiang Shi [1*]

[1]*School of Physics, Southeast University, Nanjing 211189, China*
[2]*Department of Applied Physics, The University of Tokyo, Tokyo 113-8656, Japan*
[3]*School of Physics and Physical Engineering, Qufu Normal University, Qufu 273165, China*
[4]*School of Physics and Astronomy, Shanghai Jiao Tong University Shanghai 200240, China*
[5]*Materials Genome Institute, Shanghai University. Shanghai 200444, China*
[6]*School of Physical Science and Technology, ShanghaiTech University, Shanghai 201210, China*
[7]*Institute of Plasma Physics, Chinese Academy of Sciences, Hefei, Anhui 230031, China*
[8]*State Key Laboratory of Advanced Electromagnetic Engineering and Technology, School of Electrical and Electronic Engineering, Huazhong University of Science and Technology, Wuhan, 430030, China*
[#]*Authors contributed equally to the paper*

[*]Email: sunyue@seu.edu.cn (Yue Sun); zxshi@seu.edu.cn (Zhixiang Shi)

**Abstract**

Iron telluride-selenium superconducting materials, known for their non-toxicity, ease of preparation, simple structure, and high upper critical fields, have attracted much research interest in practical application. In this work, we conducted electrical transport measurements, magneto-optical imaging, and AC magnetic susceptibility measurements on FeTe$_{0.5}$Se$_{0.5}$ superconducting long tapes fabricated via reel-to-reel pulsed laser deposition. Our transport measurements revealed a high critical current density that remains relatively stable even with increasing external magnetic fields, reaching over $1 \times 10^5$ A/cm² at 8 K and 9 T. The calculated pinning force density indicates that normal point pinning is the primary mechanism in these tapes. The magneto-optical images demonstrated that the tapes show homogeneous superconductivity and uniform distribution of critical current density. The AC magnetic susceptibility measurements also confirmed their strong flux pinning nature of withstanding high magnetic field. Based on these characteristics, FeTe$_{0.5}$Se$_{0.5}$ superconducting tapes show promising prospects for applications under high magnetic field.

## 1. Introduction

Iron-based superconductors are among the most promising candidates for practical applications, garnering significant interest [1–7]. Recent studies have particularly focused on the FeTe$_{0.5}$Se$_{0.5}$ system, especially its thin films. For instance, FeTe$_{0.5}$Se$_{0.5}$ thin films have achieved a remarkable critical current density ($J_c$) exceeding $10^6$ A/cm² under zero field at 4.2 K [8]. Research into these thin films, deposited on various substrates, reveals that the critical temperature ($T_c$) reaches its highest value of 19.1 K on CaF$_2$ substrates, compared to LaO and MgO substrates [9]. Furthermore, magneto-optical (MO) experiments have shown that FeTe$_{0.5}$Se$_{0.5}$ thin films exhibit low anisotropy [10,11], but at the same time, non-homogeneities have also been observed in these films. Additionally, a study on the effects of artificial defects introduced via proton irradiation showed that $J_c$ can be enhanced by ~ 1.6 times under zero field at 4.2 K [12].

For practical applications, superconducting tapes are often preferred over films. Recent research has focused on FeTe$_{0.5}$Se$_{0.5}$ tapes, examining factors such as the thickness of the superconducting layer. For instance, Ref. [13] reported optimal $J_c$ of 1.3 MA/cm² under 0 T at 4.2 K for 400 nm thick short tapes produced using static pulsed laser deposition (PLD), similar to thin films on crystal substrates. More recently, long tapes fabricated using reel-to-reel PLD methods were reported in Ref. [14], under 0 T at 4.2 K, with optimal $J_c$ values of 1.308 MA/cm² for a thickness of 450 nm. These long tapes are more suitable for practical applications. The same group from Ref. [14] more recently fabricated meter-long Fe(Se,Te) tapes [15]. While $J_c$ is very high at 4.2 K under zero field (~2 MA/cm²), it decreases rapidly with increasing external field ($H_a$). Under $H$ = 9 T, the $J_c$ quickly drops to 0.6 MA/cm². Clearly, this is not ideal for applications in high magnetic fields, and better tapes with lower $J_c$ decrease under high magnetic fields are needed. Additionally, Ref. [16] explored how encapsulation materials affect the transport properties of Fe(Se,Te) tapes. They found that compared to silver stabilizing layers, the combination of copper stabilizing layers and Bi$_{50}$Pb$_{25}$Sn$_{12.5}$Cd$_{12.5}$ solder more effectively enhances the transport properties of Fe(Se,Te) tapes. These studies provide valuable insights into the practical application of Fe(Se,Te) tapes. However, several critical aspects remain underexplored. Notably, there is a lack of information regarding the homogeneity of Fe(Se,Te) tapes, which directly affects their superconducting performance. Additionally, AC susceptibility of Fe(Se,Te) tapes are still not studied, which is a crucial index for evaluating superconducting performance. In addition, when it comes to high-field applications, the rate of $J_c$ decrease and the pinning force density ($F_p$) under intense magnetic fields are crucial considerations that have yet to be thoroughly comprehended.

To advance the practical application of FeTe$_{0.5}$Se$_{0.5}$ superconductors, we fabricated FeTe$_{0.5}$Se$_{0.5}$ tapes with a thickness of 400 nm using reel-to-reel PLD. We analyzed the pinning properties under high magnetic fields by measuring the transport $J_c$. MO experiments confirmed that the FeTe$_{0.5}$Se$_{0.5}$ tape exhibits good homogeneity. Furthermore, we measured the AC magnetic susceptibility (ACMS) of the tapes and examined the effects of varying alternating fields and superimposed DC fields on hysteresis AC losses.

## 2. Experimental details

FeTe$_{0.5}$Se$_{0.5}$ tapes, 50 μm thick and 10 mm wide, were fabricated on CeO$_2$-buffered IBAD metal tapes using reel-to-reel PLD [16–18]. Structural characterization was conducted with X-ray diffraction (XRD) using a Rigaku diffractometer with a Cu-Kα radiation.

DC magnetization was measured using a Magnetic Property Measurement System (MPMS-XL5, Quantum Design). MO imaging was performed with a 12-bit CCD camera (ORCA-ER, Hamamatsu) on 2.3 mm × 2 mm tapes with a 400 nm thick FeTe$_{0.5}$Se$_{0.5}$ film. The images are taken in the remanent state. The sample was cooled down to the target temperature using zero-field cooling. An external field of 800 Oe (which is large enough to fully penetrate the FeTe$_{0.5}$Se$_{0.5}$ sample, as demonstrated in the penetration study in Ref. [10]) was then applied to reach the remanent state. The measurement principle of MO can be found in the supplementary material.

Temperature-dependent resistance, voltage-current characteristics, and ACMS were measured with a Physical Property Measurement System (PPMS-9 T, Quantum Design). Electrical transport performances were obtained using a four-point probe method on 5 mm × 1.8 mm × 0.05 mm tapes with a 300 nm thick superconducting layer. ACMS was measured with zero-field cooling in $M(T)$ cycles at 0.1 K/min on tapes sized 3.5 mm × 3 mm with a 400 nm thick $FeTe_{0.5}Se_{0.5}$ film [19].

## 3. Results and discussion

Figure 1(a) shows XRD pattern of the $FeTe_{0.5}Se_{0.5}$ tape, showing observable (00$l$) peaks for $FeTe_{0.5}Se_{0.5}$ as well as substrate peaks for CeO2 (111) and Hastelloy C276. And we obtained a $c$-axis lattice parameter of 5.89 Å for $FeTe_{0.5}Se_{0.5}$ in the tapes. This also confirms the tape structure as illustrated in Figure 1(b). The temperature ($T$)-dependent magnetic susceptibility in Figure 1(c) confirms superconductivity with a $T_c$ of 15.4 K, consistent with the resistance data shown in Figure 1(d). The linear resistance above $T_c$ is attributed to the silver protective layer.

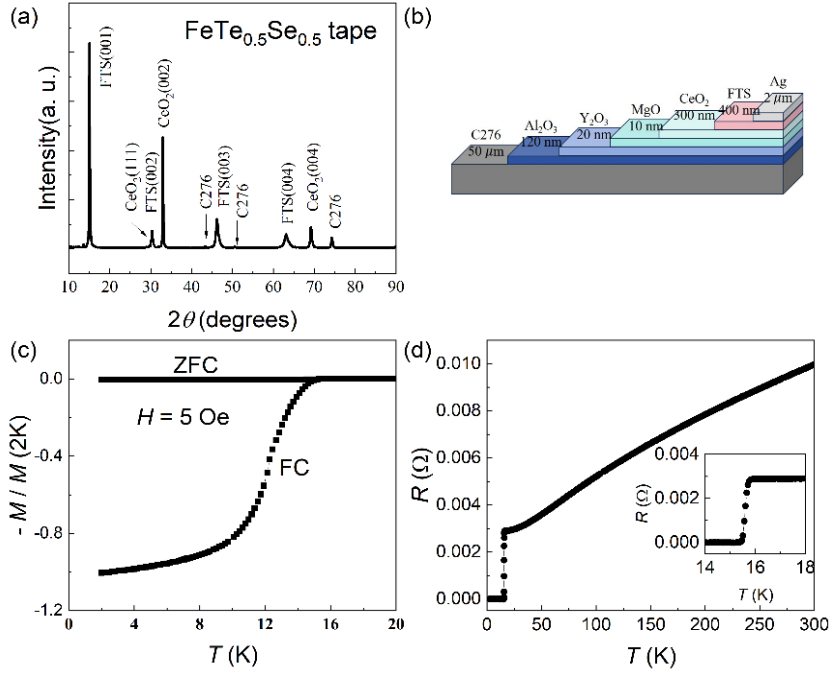

Figure 1. (a) X-ray diffraction patterns of $FeTe_{0.5}Se_{0.5}$ tape, (b) The structure of $FeTe_{0.5}Se_{0.5}$ tape. (c) Normalized magnetization versus temperature. (d) Zero-field resistance versus temperature for the $FeTe_{0.5}Se_{0.5}$ tape. The inset shows a detailed view near the $T_c$.

Figures S5(a)-S5(l) in the supplementary display voltage-current curves under 0 - 9 T ($H \parallel c$) at 3 - 14 K, obtained from transport measurements. Using the off-set method (Supplementary Figure S5(m)), the critical current ($I_c$) of the tape in Figure S5(a) at 3 K under 0 T is 1027 mA. $J_c$ derived from these measurements is depicted in Figure 2(a). $J_c$ of the $FeTe_{0.5}Se_{0.5}$ tape reaches $1.5 \times 10^5$ A/cm² under 0 T at 3 K. As the temperature and field increase, $J_c$ decreases very slowly. At 3 K and 9 T, $J_c$ is $1.3 \times 10^5$ A/cm². Even at 8 K and 9 T, $J_c$ remains larger than $1.0 \times 10^5$ A/cm². At 4.2 K, when the field increases from 0 T to 9 T, $J_c$ decreases by 30%. In contrast, previous tapes prepared by reel-to-reel PLD showed a reduction of 77% [16]. Figure 2(b) shows the distribution of $J_c$ at different temperatures and magnetic fields, with the red dashed line representing the threshold of $1.0 \times 10^5$ A/cm². The region to the left of the threshold indicates higher $J_c$ values, suggesting that this tape can be applied in a wide range of magnetic fields at temperatures below 10K.

To investigate the pinning mechanism of the $FeTe_{0.5}Se_{0.5}$ tape, $F_p$ at various temperatures was derived using the formula $F_p = J_c \times H$, as illustrated in Figure 2(c). The $F_p$ increases with the magnetic field and has not yet saturated within the range of magnetic fields used in our experiments at temperatures below 12 K. Evidently, the peak of $F_p$ is observed only when temperature near the $T_c$. The Dew-Hughes model [20] suggests that if a single pinning mechanism prevails within a specific temperature region, the normalized pinning force density $f_p$ can be scaled by $h$, where $f_p = F_p / F_{pmax}$, $F_p = J_c \times H$, $h = H / H_{c2}$, and $F_{pmax}$ represents the maximum pinning force density. For high-temperature superconductors (HTS) with a large upper critical field $H_{c2}$, the Higuchi method is often used. Specifically, scaling is performed using $h = H / H_{max}$, where $H_{max}$ is the magnetic field corresponding to $F_{pmax}$ [21–24]. $f_p$ can be depicted as:

(a) $\Delta\kappa$ pinning ($\kappa$ is the Ginzburg-Landau parameter),
$$f_p = 3h^2(1-2h/3), \quad (1)$$
(b) Normal point pinning,
$$f_p = 9h(1-h/3)^2/4, \quad (2)$$
(c) Surface pinning,
$$f_p = 25h^{1/2}(1-h/5)^2/16. \quad (3)$$

The inset in Figure 2(c) shows the relationship between $F_p / F_{pmax}$ and $H / H_{max}$ at 12 K and 13 K. It is clear that the data from both temperatures can be scaled onto a single master curve, indicating that the pinning mechanism is the same at these temperatures. Furthermore, the $f_p$ - $h$ curve matches well with equation (2) from the Higuchi method, suggesting that normal point pinning is the dominant mechanism.

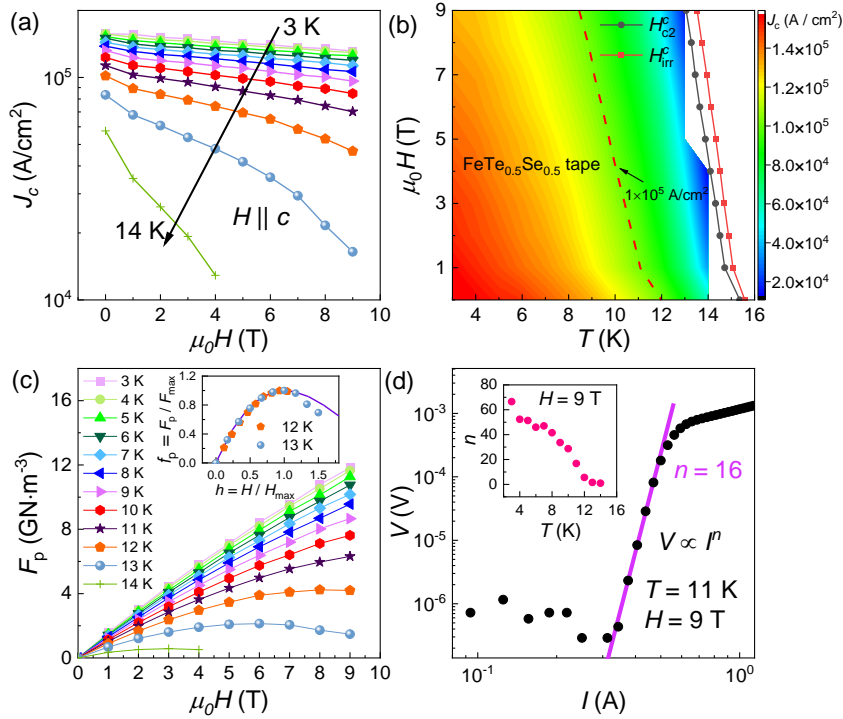

Figure 2. (a) The relationship between $J_c$ and external magnetic field for FeTe$_{0.5}$Se$_{0.5}$ tape at different temperatures. The temperature labels are the same as in (c). (b) Phase diagram of $J_c$ as a function of temperature and magnetic field. The red dashed line represents the boundary at $1.0 \times 10^5$ A/cm$^2$. The black solid circles indicate $H_{c2}$ when $H \parallel c$, while the red solid squares denote $H_{irr}$ under the same conditions. (c) $F_p$ versus magnetic field, with the inset showing the variation of $F_p / F_{pmax}$ at lower magnetic fields for 12 K and 13 K. (d) The voltage versus current at 11 K under 9 T, with the inset illustrating the n-index schematic at different temperatures.

To quantitatively investigate the current-carrying capability of the tape near $I_c$, the $I$-$V$ curve was fitted using $V \propto I^n$. The $n$ value describes the homogeneity of the superconducting material, the thermally activated depinning behavior, and can be used to evaluate the pinning mechanisms and pinning forces. At lower temperatures, the $n$ value is primarily influenced by homogeneity. As temperature and magnetic field increase, the influence of homogeneity decreases while the impact of flux creep becomes more significant [25-27].

In HTS applications, a higher $n$ value is crucial for superconducting coils, as it helps reduce losses when operating near $J_c$ and is essential for magnets running in persistent mode [28]. As depicted in Figure 2(d), at $T = 11$ K and 9 T, the superconducting transition region aligns well with $V \propto I^{16}$. Similarly, the inset shows $n$ values from 3 K to 14 K under 9 T, with $n$ reaching 52 at 4 K. The $n$ value for YBCO tape is typically around ~30 (at 77 K and 0 T) [29]. Under 10 T at 4.2 K, the $n$ value for Nb$_3$Sn is ~30 [30]. At low temperatures, the $n$ value of FeTe$_{0.5}$Se$_{0.5}$ tapes is higher compared to Nb$_3$Sn, indicating excellent homogeneity. The higher $n$ value near $T_c$ suggests strong pinning potential and low flux creep in the tape.

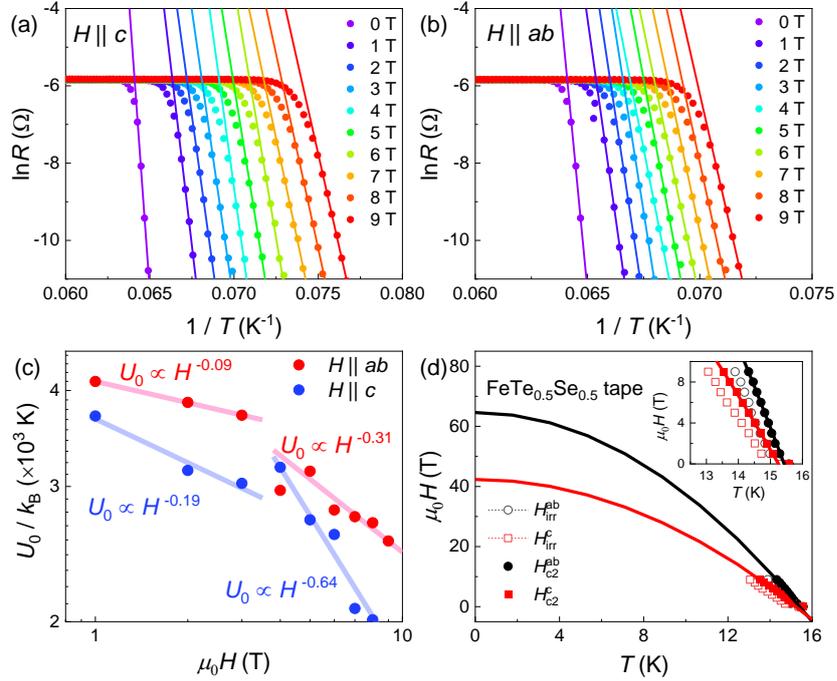

Figure 3. The relationship between $\ln R$ and $1/T$ under (a) $H \parallel c$ and (b) $H \parallel ab$, where the straight line represents Arrhenius equation fit. (c) $U_0$ obtained from fitting the Arrhenius equation, along with the fitting to $U_0 \propto H^\alpha$. (d) $H_{c2}$ versus temperature, with the solid line showing G-L fitting curve. The inset provides a detailed view near $T_c$.

To examine how external fields affect the performance of the tapes, we measured the resistance versus temperature under different fields, as shown in Figure S6. We obtained the relationship between $\ln R$ and $1/T$ as shown in Figures 3(a) and 3(b). Using the Arrhenius formula $\ln\rho_0(H) = \ln\rho_0 - U_0/k_B T_c$, ($U_0$ is the pinning potential and $k_B$ is the Boltzmann constant) for fitting [31-36], we determined the relationship between the pinning potential $U_0(H)$ and the field in Figure 3(c). The figure shows that $U_0$ of the external field (at logarithmic scale) presents two different linear regions for $H \parallel c$ and $H \parallel ab$. Under $H \parallel c$, $U_0(H) \propto H^{-0.19}$ for $H < 4$ T and $U_0(H) \propto H^{-0.64}$ for $H > 4$ T. This demonstrates that the tape exhibits different predominant pinning types under low and high fields, with single-flux pinning being predominant at low fields and collective pinning under high fields. Specifically, weak pinning is observed at high fields and strong pinning at low fields. Conversely, for $H \parallel ab$, the exponent $\alpha$ remains below 0.5 from 0 T to 9 T, signifying a strong pinning effect. Nonetheless, the $\alpha$ index shows a sharp variation under $H = 4$ T.

Additionally, to determine the relationship between $H_{c2}$ and the irreversibility field ($H_{irr}$) with temperature, we calculate the first derivative of the resistance versus temperature. This analysis yields the $H_{c2}$ at the peak temperature $T_p$ and the $H_{irr}$ at the ending point of superconducting transition $T_p^*$ on the left side of the peak (Supplementary Material Figures S6(c) and S6(d)). According to the G-L formula [37,38], $H_{c2}$ was fitted as shown in Figure 3(d). The tape's $\mu_0 H_{c2}^{ab}(0)$ is 65 T and $\mu_0 H_{c2}^{c}(0)$ is 43 T. As illustrated in the inset, the $H_{irr}$ in any direction is also significant (~9 T at 13 K). The $H_{c2}$ and $H_{irr}$ are relatively large, with an out-of-plane anisotropy of only 1.67. The coherence length $\xi_{ab}(0\ K)$ and $\xi_c(0\ K)$ could be computed using the formula: $\xi_{ab}(0\ K) = (\Phi_0/(2\pi H_{c2}^c))^{-1/2}$, $\xi_c(0\ K) = \Phi_0/(2\pi\xi_{ab}(0\ K)\mu_0 H_{c2}^{ab}(0\ K))$, where $\Phi_0$ is the magnetic flux quantum. The results show that $\xi_{ab}(0\ K) = 2.76$ nm and $\xi_c(0\ K) = 1.83$ nm. Therefore, introducing defects with a size (of ~5.5 nm) approximately twice that of $\xi_{ab}(0\ K)$ is beneficial for optimizing the pinning performance of the tape. Additionally, based on the assessment that the pinning type of the tape is normal-point pinning, the peak of $F_p$ at 4.2 K is found to be ~ 20 T ($H \parallel ab$). This indicates that the tape is well-suited for applications at high magnetic fields.

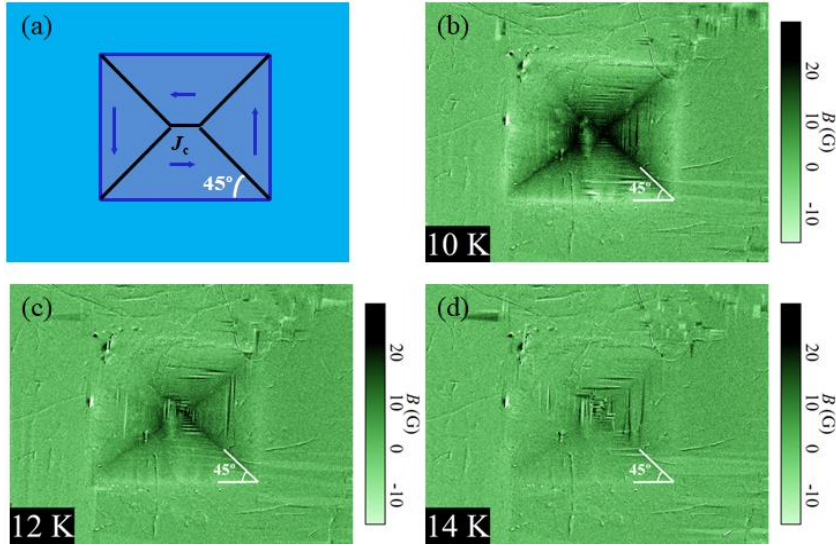

Figure 4. (a) Schematic illustration of the $J_c$ distribution in a homogeneous rectangular superconductor. MO images of FeTe$_{0.5}$Se$_{0.5}$ tape in the remanent state at (b) $T = 10$ K, (c) $T = 12$ K, and (d) $T = 14$ K. More images for $T = 11$ K, 13 K, and 15 K can be found in supplementary materials.

Homogeneity is one of the important parameters for superconducting tapes. In this study, we conducted MO imaging experiments to directly reveal the homogeneity of FeTe$_{0.5}$Se$_{0.5}$ tape. In an ideal homogeneous superconductor with a regular shape, when it reaches the critical state, the current should flow along the edge, resulting in double Y-shaped lines, as shown in Figure 4(a). In other words, the angle between the edge and the Y-shaped line forms 45º if the current flows homogeneously around the sample surface. As shown in Figures. 4(b)-(d), a clear double-Y shaped current discontinuity line (angle ~ 45°) can be observed, indicating a uniform current flow on the tape surface and a uniform in-plane $J_c$. To further evaluate uniformity, we focused on data collected near the $T_c$. For our magneto-optical MO images, the angle indicated by the white line remains nearly unchanged with increasing temperature, confirming that this tape exhibits excellent homogeneity up to $T_c$. Compared to previous MO observations on FeTe$_{0.5}$Se$_{0.5}$ thin films [10,11], our MO images demonstrate significantly better homogeneity. In Ref. [10,11], at low temperatures, multi-branched structures are observed, while an island-like structures appear as the temperature approaches $T_c$. These irregular shapes result from the inhomogeneity of the tape. However, neither the multi-branched structures nor the island-like structures were observed in present study, demonstrating the excellent homogeneity of our tape. Additionally, we used a larger tape (~2.3 mm × 2 mm) for MO observations rather than the small film used in Refs. [10,11] (Width and length less than 0.6 mm), and our larger tape shows better homogeneity. We also estimated the $J_c$ using the formula [10] of

$$\Delta B = \frac{8 J_c \cdot 2d}{c} \int_0^a \frac{\omega^2}{\sqrt{z_g^2 + 2\omega^2 \left( z_b^2 + \omega^2 \right)}} d\omega, \quad (4)$$

where $\Delta B$ represents the field trapped in the tape, $2a$ denotes the length/width, $2d$ is the superconducting layer's thickness, $z_g$ is the gap between the tape surface of the tape and the MO indicator, and $c$ is the velocity of light, equal to 10 in CGS units. Using this formula, $J_c$ is ~ 300 kA/cm² under 0 T at 10 K, which aligns well with the value estimated from our $I$-$V$ measurements.

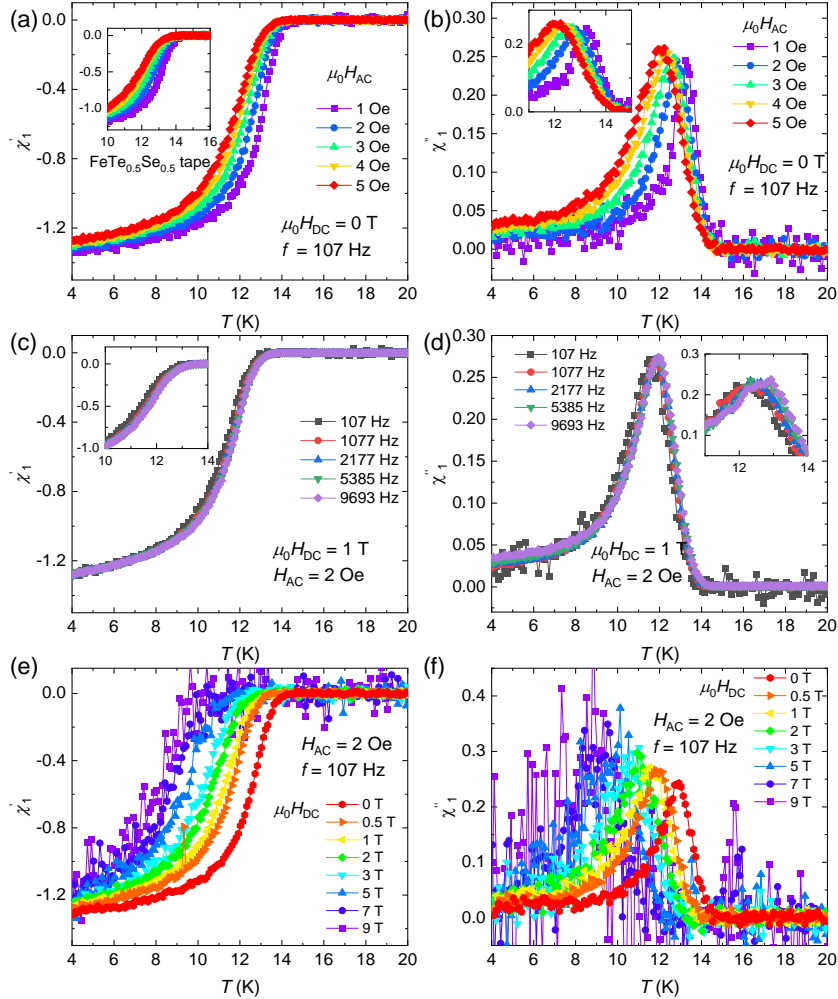

Figure 5. ACMS vs $T$: (a) real part and (b) imaginary part for different $H_{AC}$ under a 0 T DC field at 107 Hz. ACMS vs $T$: (c) real part and (d) imaginary part for different AC field frequencies under a 1 T DC field with a 2 Oe AC field amplitude at 107 Hz. ACMS vs $T$: (e) real part and (f) imaginary part for different DC fields with a 2 Oe AC field at 107 Hz.

As final characterizations of our tape, we conducted AC susceptibility measurements. **Firstly**, the real and imaginary parts of ACMS under $H_{DC}$ = 0 T and $H_{AC}$ with frequence of 107 Hz ($f_{AC}$ = 107 Hz) are depicted in Figures 5(a-b). Near $T_c$, the magnetic field begins to penetrate the tape. $\chi'_1$ reveals the tape's diamagnetic response, and the starting point of its negative signal can also be used to determine $T_c$. The changes in $\chi''_1$ reflect the magnetic field penetration effect. The peak of $\chi''_1$ typically corresponds to the temperatures at which the field penetrates the tape. This means that the penetration field $H_p$ at this temperature point is equal to $H_{AC}$. At this temperature, the response of the magnetic flux vortices to AC field is maximized, causing $\chi''_1$ to reach its peak value. According to the Bean model, when the tape thickness remains constant, the penetration field is directly proportional to $J_c$. Therefore, as $H_{AC}$ increases, the $\chi''_1$ peak shifts to lower temperature with higher $J_c$ [39], as shown in Figure 5(b). **Secondly**, the ACMS under $H_{DC}$ = 1 T and $H_{AC}$ = 2 Oe with different frequencies is shown in Figures 5(c-d). In the absence of flux creep, the peak position should remain unchanged with frequency variations. If flux creep occurs, the peak position shifts as the frequency changes. In our case, the peak position of $\chi''_1$ shifts slowly towards higher temperatures as the frequency increases, reflecting the small effect of flux creep on the AC susceptibility $\chi''$. The weak flux creep effect indicates a strong pinning potential in the tape, consistent with the large pinning potential values in Fig.3(c). **Finally**, the ACMS under different $H_{DC}$ with $H_{AC}$ = 2 Oe and $f_{AC}$ = 107 Hz is shown in Figures 5(e) and 5(f), respectively. Evidently, with the increase in $H_{DC}$, the starting point of the negative signal in $\chi'_1$ shifts to lower temperatures, indicating that the external field suppresses the superconducting state.

In AC external field, the magnetization of the tape also exhibits periodic variations. Within a single cycle, the hysteresis AC loss ($Q$) per unit volume of the tape corresponds to the area enclosed by a hysteresis loop, which can be accurately determined using the following formula [40,41]:

$$Q = \oint \mu_0 H dM = \pi \mu_0 H_{ac}^2 \chi''. \tag{5}$$

In Figure 6(a), the relationship between the $Q$ and the $H_{AC}$ under $H_{DC}$ = 1 T and $f_{AC}$ = 107 Hz can be observed. By utilizing the $Q \propto H^\alpha$ fitting approach for the experimental data points, we can derive the exponent $\alpha$ across various temperatures and AC field amplitudes. This exponent serves as an indicator of the changing rate in $Q$ influenced by $H_{AC}$. According to the Bean model, before the AC magnetic field penetrates the sample, both the extent of field penetration and $H_{AC}$ contribute to the increase in $Q$. At this stage, $\alpha$ should be 3. Once the magnetic field has fully penetrated the tape, $Q$ is influenced only by $H_{AC}$, and at this point, $\alpha$ equals 1. According to Equation (5), when $\chi''$ remains nearly constant, $Q$ is proportional to the square of $H_{AC}$, meaning $\alpha$ should be 2. This is consistent with Figures S8(a) and S9(b) (Supplementary Materials). Therefore, the observed $\alpha$ value of 2.88 at 10 K indicates that the AC magnetic field has not yet penetrated the tape. At 13 K, with an $\alpha$ value is ~1.05, suggesting that the AC magnetic field has fully penetrated the tape.

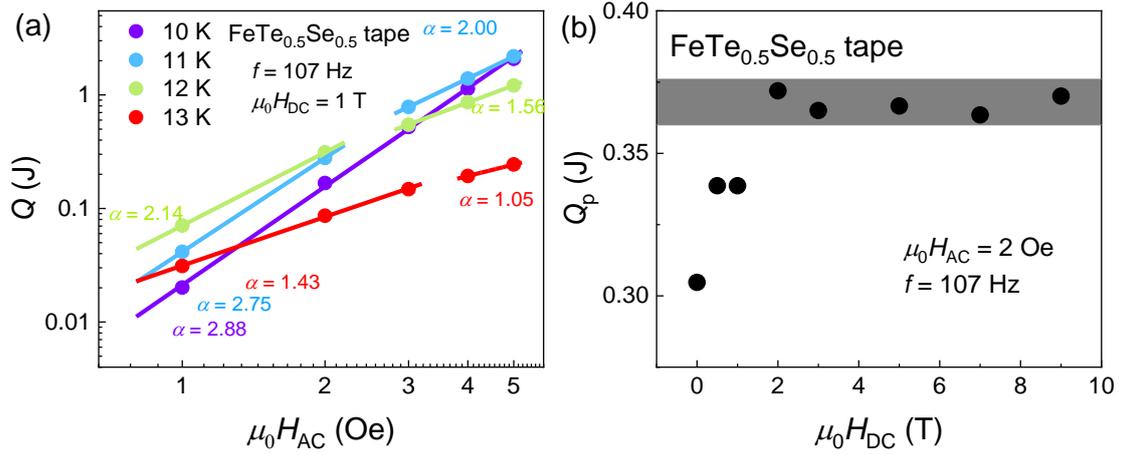

Figure 6. (a) Illustrates the correlation between $Q$ and $H_{AC}$, when $H_{DC}$ = 1T and $f_{AC}$ = 107 Hz. The straight line depicted represents the fitted curve of $Q \propto H^{\alpha}$. (b) The relationship between $Q_p$ and $H_{DC}$.

We define the hysteretic AC loss at the peak of $\chi''$ as $Q_p$. As shown in Figure 6(b), the maximum hysteresis AC loss ($Q_p$) increases significantly when $H_{DC}$ is slightly increased from zero. The $Q_p$ is related to both the tape dimensions and the $J_c$. At 9-14 K and under 2-9 T, $Q_p$ shows minimal variation, indicating the field tolerance characteristics of $J_c$. Using the formula $J_c = H_{ac}/2d$ (2d is the superconducting layer's thickness) [42], the $J_c$ of the tape can be estimated as $3.2 \times 10^5$ A/cm$^2$, which is consistent with the results from MO and transport measurements. In contrast, MgB$_{2-x}$C$_x$ is significantly more affected by the DC field, and its $Q_p$ is only about 1/3 of that of FeTe$_{0.5}$Se$_{0.5}$ tapes (under an AC field of 5 Oe and 5111 Hz) [43]. This indicates that the pinning in FTS tapes is stronger than that in MgB$_2$. Since $Q_p$ is dependent on dimensions, we plan to etch the tapes into multiple parallel strands to reduce the total hysteresis AC loss in the tapes which is expected to reduce the loss to 10% of its current value.

## 4. Conclusion

In summary, our systematic study reveals that high-quality FeTe$_{0.5}$Se$_{0.5}$ tapes offer several advantages. They exhibit high $H_{c2}$ and $H_{irr}$, along with a relatively low out-of-plane anisotropy of 1.67. The $J_c$ decreases very slowly with the increase of temperature and field. MO images confirm the excellent homogeneity of the tapes. These characteristics suggest that FeTe$_{0.5}$Se$_{0.5}$ tapes are well-suited for applications in low-temperature and high-field environments. However, further efforts are needed to reduce hysteresis AC losses to optimize their performance.

**Acknowledgements**

This work was supported by the National Key R&D Program of China (Grant No. 2018YFA0704300) and the Fundamental Research Funds for the Central Universities 2242024k30029.

**References**

[1] H. Hosono, A. Yamamoto, H. Hiramatsu, and Y. Ma, *Recent Advances in Iron-Based Superconductors toward Applications*, Materials Today **21**, 3 (2018). https://doi.org/10.1016/j.mattod.2017.09.006

[2] T. Tamegai, T. Taen, H. Yagyuda, Y. Tsuchiya, S. Mohan, T. Taniguchi, Y. Nakajima, S. Okayasu, M. Sasase, H. K., T. Murakami, T. Kambara, and Y. Kanai, *Effects of particle irradiations on vortex states in iron-based superconductors*, Supercond. Sci. Technol. **25**, 084008 (2012). https://doi.org/10.1088/0953-2048/25/8/084008

[3] Y. Sun, Z. Shi, and T. Tamegai, Review of annealing effects and superconductivity in $Fe_{1+y}Te_{1-x}Se_x$ superconductors, Supercond. Sci. Technol. **32**, 103001 (2019). https://doi.org/10.1088/1361-6668/ab30c2

[4] A. Iyo, K. Kawashima, T. Kinjo, T. Nishio, S. Ishida, H. Fujihisa, Y. Gotoh, K. Kihou, H. Eisaki, and Y. Yoshida, *New-Structure-Type Fe-Based Superconductors: $CaAFe_4As_4$ (A = K, Rb, Cs) and $SrAFe_4As_4$ (A = Rb, Cs)*, J. Am. Chem. Soc. **138**, 10 (2016). https://doi.org/10.1021/jacs.5b12571

[5] F.-C. Hsu *et al.*, *Superconductivity in the PbO-Type Structure α-FeSe*, Proceedings of the National Academy of Sciences **105**, 38 (2008). https://doi.org/10.1073/pnas.0807325105

[6] T. Shibauchi, T. Hanaguri, and Y. Matsuda, *Exotic Superconducting States in FeSe-based Materials*, J. Phys. Soc. Jpn. **89**, 102002 (2020). https://doi.org/10.7566/JPSJ.89.102002

[7] W. Zhou, X. Xing, W. Wu, H. Zhao, and Z. Shi, *Second magnetization peak effect, vortex dynamics and flux pinning in 112-type superconductor $Ca_{0.8}La_{0.2}Fe_{1-x}Co_xAs_2$*, Sci. Rep. **6**, 22278 (2016).

[8] W. Si, S. J. Han, X. Shi, S. N. Ehrlich, J. Jaroszynski, A. Goyal, and Q. Li, *High Current Superconductivity in $FeSe_{0.5}Te_{0.5}$-Coated Conductors at 30 Tesla*, Nat. Commun. **4**, 1347 (2013). https://doi.org/10.1038/ncomms2337

[9] F. Yuan *et al.*, *Influence of Substrate Type on Transport Properties of Superconducting $FeSe_{0.5}Te_{0.5}$ Thin Films*, Supercond. Sci. Technol. **28**, 065005 (2015). https://doi.org/10.1088/0953-2048/28/6/065005

[10] Y. Sun, Y. Tsuchiya, S. Pyon, T. Tamegai, C. Zhang, T. Ozaki, and Q. Li, *Magneto-Optical Characterizations of $FeTe_{0.5}Se_{0.5}$ Thin Films with Critical Current Density over 1 MA $Cm^{-2}$*, Supercond. Sci. Technol. **28**, 015010 (2014). https://doi.org/10.1088/0953-2048/28/1/015010

[11] P. Yuan, Z. Xu, Y. Ma, Y. Sun, and T. Tamegai, *Angular-Dependent Vortex Pinning Mechanism and Magneto-Optical Characterizations of $FeSe_{0.5}Te_{0.5}$ Thin Films Grown on $CaF_2$ Substrates*, Supercond. Sci. Technol. **29**, 035013 (2016). https://doi.org/10.1088/0953-2048/29/3/035013

[12] T. Ozaki, L. Wu, C. Zhang, J. Jaroszynski, W. Si, J. Zhou, Y. Zhu, and Q. Li, *A Route for a Strong Increase of Critical Current in Nanostrained Iron-Based Superconductors*, Nat Commun **7**, 1 (2016). https://doi.org/10.1038/ncomms13036

[13] J. Song *et al.*, *Critical Role Played by Interface Engineering in Weakening Thickness Dependence of Superconducting and Structural Properties of $FeSe_{0.5}Te_{0.5}$-Coated Conductors*, ACS Appl. Mater. Interfaces **15**, 26215 (2023). https://doi.org/10.1021/acsami.3c04531

[14] J. Ye, S. Mou, R. Zhu, L. Liu, and Y. Li, *Evolution of Superconductivity Dependence on Substrate Temperature with Thickness of Fe(Se,Te) Coated Conductors Deposited on Metal Tapes*, Journal of Applied Physics **132**, 183910 (2022). https://doi.org/10.1063/5.0122319

[15] L. Liu *et al.*, *Fabrication of Meter-Long Class Fe(Se,Te)-Coated Conductors with High

*Superconducting Performance*, Advanced Engineering Materials **25**, 2201536 (2023). https://doi.org/10.1002/adem.202201536

[16] X. Liu *et al.*, *Study on the Transport Current of FeSe$_{0.5}$Te$_{0.5}$ Coated Conductor with Different Stabilizing Layers and Solders*, Cryogenics **133**, 103711 (2023). https://doi.org/10.1016/j.cryogenics.2023.103711

[17] Y. Li, L. Liu, and X. Wu, *Fabrication of Long REBCO Coated Conductors by PLD Process in China*, Physica C: Superconductivity and Its Applications **518**, 51 (2015). https://doi.org/10.1016/j.physc.2015.03.020

[18] S. Wei *et al.*, *First Performance Test of FeSe0.5Te0.5-Coated Conductor Coil under High Magnetic Fields*, Supercond. Sci. Technol. **36**, 4 (2023). https://doi.org/10.1088/1361-6668/acba4d

[19] A. Galluzzi, K. Buchkov, V. Tomov, E. Nazarova, A. Leo, G. Grimaldi, S. Pace, and M. Polichetti, *Mixed State Properties Analysis in AC Magnetic Field of Strong Pinning Fe(Se,Te) Single Crystal*, Supercond. Sci. Technol. **33**, 094006 (2020). https://doi.org/10.1088/1361-6668/aba354

[20] D. Dew-Hughes, *Flux Pinning Mechanisms in Type II Superconductors*, The Philosophical Magazine: A Journal of Theoretical Experimental and Applied Physics **30**, 293 (1974). https://doi.org/10.1080/14786439808206556

[21] M. Li, L. Chen, W.-L. You, J. Ge, and J. Zhang, *Giant Increase of Critical Current Density and Vortex Pinning in Mn Doped K$_x$Fe$_{2-y}$Se$_2$ Single Crystals*, Applied Physics Letters **105**, 192602 (2014). https://doi.org/10.1063/1.4901902

[22] T. Higuchi, S. I. Yoo, and M. Murakami, *Comparative Study of Critical Current Densities and Flux Pinning among a Flux-Grown NdBa2Cu3Oy Single Crystal, Melt-Textured Nd-Ba-Cu-O, and Y-Ba-Cu-O Bulks*, Phys. Rev. B **59**, 1514 (1999). https://doi.org/10.1103/PhysRevB.59.1514

[23] L. Klein, E. R. Yacoby, Y. Yeshurun, A. Erb, G. Müller-Vogt, V. Breit, and H. Wühl, *Peak Effect and Scaling of Irreversible Properties in Untwinned Y-Ba-Cu-O Crystals*, Phys. Rev. B **49**, 4403 (1994). https://doi.org/10.1103/PhysRevB.49.4403

[24] Y. Pan, N. Zhou, B. Lin, J. Wang, Z. Zhu, W. Zhou, Y. Sun, and Z. Shi, *Anisotropic Critical Current Density and Flux Pinning Mechanism of Fe$_{1+y}$Te$_{0.6}$Se$_{0.4}$ Single Crystals*, Supercond. Sci. Technol. **35**, 015002 (2021). https://doi.org/10.1088/1361-6668/ac3632

[25] M. Parizh, Y. Lvovsky, and M. Sumption, *Conductors for commercial MRI magnets beyond NbTi: Requirements and challenges*, Supercond. Sci. Technol. **30**, 014007 (2017). https://doi.org/10.1088/0953-2048/30/1/014007

[26] Z. Wang, Z. Chen, Y. Zhou, Z. Duan, and W. Wang, *The effect of temperature on Jc and n-value of Bi(2223) tapes*, Cryogenics **40**, 681 (2000). https://doi.org/10.1016/S0011-2275(00)00068-0

[27] D. Gajda, A. J. Zaleski, M. Babij, M. A. Rindfleisch, *Indication of the measuring method for accurately determining the critical current and n value in superconducting wires and tapes used in superconducting coils*, Journal of Applied Physics **136**, 053901 (2024). https://doi.org/10.1063/5.0219306

[28] M. Chudy, Z. Zhong, M. Eisterer, and T. Coombs, *n-values of commercial YBCO tapes before and after irradiation by fast neutrons*, Supercond. Sci. Technol. **28**, 035008 (2015). https://doi.org/10.1088/0953-2048/28/3/035008

[29] J. Ogawa, Y. Sawai, H. Nakayama, O. Tsukamoto, and D. Miyagi, *n Value and $J_c$ Distribution Dependence of AC Transport Current Losses in HTS Conductors*, Physica C: Superconductivity **401**, 171 (2004). https://doi.org/10.1016/j.physc.2003.09.031

[30] M. Liang, P. X. Zhang, X. D. Tang, J. S. Li, C. G. Li, K. Li, M. Yang, C. J. Xiao, and L. Zhou, *Effects of Heat Treatments on the Nb3Sn Composite Strands*, Materials Science Forum **546–549**, 2023 (2007). https://doi.org/10.4028/www.scientific.net/MSF.546-549.2023

[31] X. G. Qiu, B. R. Zhao, S. Q. Guo, J. R. Zhang, L. Li, F. Ichikawa, T. Nishizaki, T. Fukami, Y. Horie, and T. Aomine, *Thermally Activated Flux Dissipation in C-Axis-Oriented $YBa_2Cu_3O_7/PrBa_2Cu_3O_7$ Multilayers*, Phys. Rev. B **47**, 14519 (1993). https://doi.org/10.1103/PhysRevB.47.14519

[32] G. Blatter, M. V. Feigel'man, V. B. Geshkenbein, A. I. Larkin, and V. M. Vinokur, *Vortices in High-Temperature Superconductors*, Rev. Mod. Phys. **66**, 1125 (1994). https://doi.org/10.1103/RevModPhys.66.1125

[33] T. T. M. Palstra, B. Batlogg, R. B. van Dover, L. F. Schneemeyer, and J. V. Waszczak, *Dissipative Flux Motion in High-Temperature Superconductors*, Phys. Rev. B **41**, 6621 (1990). https://doi.org/10.1103/PhysRevB.41.6621

[34] Y. Yeshurun and A. P. Malozemoff, *Giant Flux Creep and Irreversibility in an Y-Ba-Cu-O Crystal: An Alternative to the Superconducting-Glass Model*, Phys. Rev. Lett. **60**, 2202 (1988). https://doi.org/10.1103/PhysRevLett.60.2202

[35] M. Tinkham, *Resistive Transition of High-Temperature Superconductors*, Phys. Rev. Lett. **61**, 1658 (1988). https://doi.org/10.1103/PhysRevLett.61.1658

[36] W. J. Choi, Y. I. Seo, D. Ahmad, and Y. S. Kwon, *Thermal Activation Energy of 3D Vortex Matter in $NaFe_{1-x}Co_xAs$ (x = 0.01, 0.03 and 0.07) Single Crystals*, Sci Rep **7**, 10900 (2017). https://doi.org/10.1038/s41598-017-11371-1

[37] W. Zhong, H. Zhang, E. Karaca, J. Zhou, S. Kawaguchi, H. Kadobayashi, X. Yu, D. Errandonea, B. Yue, and F. Hong, *Pressure-Sensitive Multiple Superconducting Phases and Their Structural Origin in Van Der Waals $HfS_2$ Up to 160 GPa*, Phys. Rev. Lett. **133**, 066001 (2024). https://doi.org/10.1103/PhysRevLett.133.066001

[38] K.-H. Müller, G. Fuchs, A. Handstein, K. Nenkov, V. N. Narozhnyi, and D. Eckert, *The Upper Critical Field in Superconducting $MgB_2$*, Journal of Alloys and Compounds **322**, L10 (2001). https://doi.org/10.1016/S0925-8388(01)01197-5

[39] C. P. Bean, *Magnetization of High-Field Superconductors*, Rev. Mod. Phys. **36**, 31 (1964). https://doi.org/10.1103/RevModPhys.36.31

[40] M. V. Feigel'man, V. B. Geshkenbein, A. I. Larkin, and V. M. Vinokur, *Theory of Collective Flux Creep*, Phys. Rev. Lett. **63**, 2303 (1989). https://doi.org/10.1103/PhysRevLett.63.2303

[41] W. T. Norris, *Calculation of Hysteresis Losses in Hard Superconductors Carrying Ac: Isolated Conductors and Edges of Thin Sheets*, J. Phys. D: Appl. Phys. **3**, 489 (1970). https://doi.org/10.1088/0022-3727/3/4/308

[42] Li-Xin Gao, Teng Wang, Qi-Ling Xiao, Wen-Lai Lu, Fei Chen, Gang Mu and Jun-Yi Ge, *Investigation of the flux dynamics in $KCa_2Fe_4As_4F_2$ single crystal by ac susceptibility measurements*, Supercond. Sci. Technol. **35**, 055013 (2022). https://doi.org/10.1088/1361-6668/ac5f14

[43] I. Amvrazi and M. Pissas, *Angular Dependence of the Peak Effect in $MgB_{2-x}C_x$*, Physica C: Superconductivity and Its Applications **503**, 42 (2014).

# Supplementary Material

# Critical Current Density and AC Magnetic Susceptibility of High-quality FeTe$_{0.5}$Se$_{0.5}$ Superconducting Tapes

Xin Zhou [1, #], Wenjie Li [1, 2, #], Qiang Hou[1], Wei Wei [1], Wenhui Liu [1], Ke Wang[1], Xiangzhuo Xing [3], Linfei Liu [4], Jun-Yi Ge[5], Yanpeng Qi[6], Huajun Liu[7], Li Ren[8], Tsuyoshi Tamegai[2], Yue Sun [1*] and Zhixiang Shi [1*]

[1]*School of Physics, Southeast University, Nanjing 211189, China*
[2]*Department of Applied Physics, The University of Tokyo, Tokyo 113-8656, Japan*
[3]*School of Physics and Physical Engineering, Qufu Normal University, Qufu 273165, China*
[4]*School of Physics and Astronomy, Shanghai Jiao Tong University Shanghai 200240, China*
[5]*Materials Genome Institute, Shanghai University. Shanghai 200444, China*
[6]*School of Physical Science and Technology, ShanghaiTech University, Shanghai 201210, China*
[7]*Institute of Plasma Physics, Chinese Academy of Sciences, Hefei, Anhui 230031, China*
[8]*State Key Laboratory of Advanced Electromagnetic Engineering and Technology, School of Electrical and Electronic Engineering, Huazhong University of Science and Technology, Wuhan, 430030, China*

[#]*Authors contributed equally to the paper*

[*]Email: sunyue@seu.edu.cn (Yue Sun); zxshi@seu.edu.cn (Zhixiang Shi)

# 1. Experimental Methods

## 1.1 *I-V*

First, a 3 mm × 3 mm × 0.5 mm glass substrate was adhered to the AC transport puck. Next, a tape measuring 8 mm × 2.5 mm × 300 nm (thickness of the superconducting layer) was attached to the substrate (the long strip shape with a thickness of 300 nm was designed to limit the maximum $I_c$ to between 0 and 2 A). The four-probe method was used to connect the sample to the $I^+$, $I^-$, $V^+$, and $V^-$ electrodes on the puck, where $I^+$ and $I^-$ were copper wires with a diameter of 200 μm, and $V^+$ and $V^-$ were gold wires with a diameter of 20 μm. A sine wave AC current with a maximum value of 1.2 A was applied to the sample using the PPMS for one-quarter of a cycle.

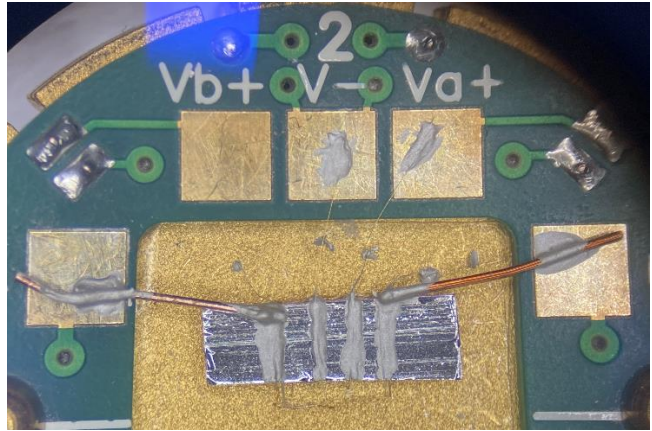

Figure S1. Schematic of the four-probe method for *I-V* measurements. The voltage electrodes are connected with gold wires, while the current electrodes are connected with copper wires.

## 1.2 *R-T*

First, a 3 mm × 3 mm × 0.5 mm glass substrate is adhered to the PPMS rotating puck, and then a tape measuring 3 mm × 1 mm × 300 nm (thickness of the superconducting layer) was attached to the substrate. The four-probe method was used to connect the sample to the $I^+$, $I^-$, $V^+$, and $V^-$ electrodes on the puck. Measurements were initially taken at an angle of 0° ($H \parallel c$) to obtain the *R-T* relationship curves under different external fields ($H$ = 0, 1, 2, 3, 4, 5, 6, 7, 8, and 9 T). Subsequently, measurements were performed at an angle of 90° ($H \parallel ab$) to obtain the *R-T* curves under various external fields ($H$ = 0, 1, 2, 3, 4, 5, 6, 7, 8, and 9 T).

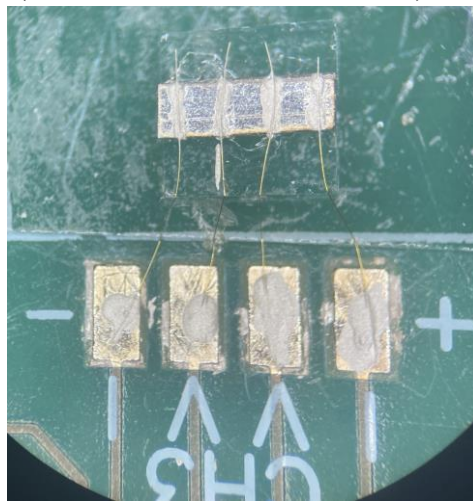

Figure S2. Four-probe method for *R-T* measurements at various angles.

## 1.3 Magnetic optical measurement

The MO measurements are based on the Faraday Effect. The basic principle involves using optical microscopy to observe brightness changes caused by the magnetic field distribution of a material in close contact with a garnet film. As illustrated in Fig. S3, light emitted from a polarized light source is directed onto the reflecting surface of the garnet film, with the sample placed directly beneath. When the light passes through the garnet and reflects from the bottom, its polarization plane is altered due to the Faraday Effect. These changes are captured by a CCD camera. As long as the garnet and the sample are in direct contact with no gaps, the magnetic field distribution through the garnet film can accurately represent the magnetic field distribution on the sample's surface. For the measurement, we first obtain reference data for the relationship between the magnetic field and light intensity at temperatures slightly above $T_c$. For the remanent measurement, the superconductor is first cooled using zero-field cooling, after which an external field of 800 Oe is applied to fully penetrate the superconductor, and then the field is set to zero to achieve the remanent state. Following the measurement, the system temperature is raised to 20 K, and the zero-field cooling process is repeated for measurements at each temperature.

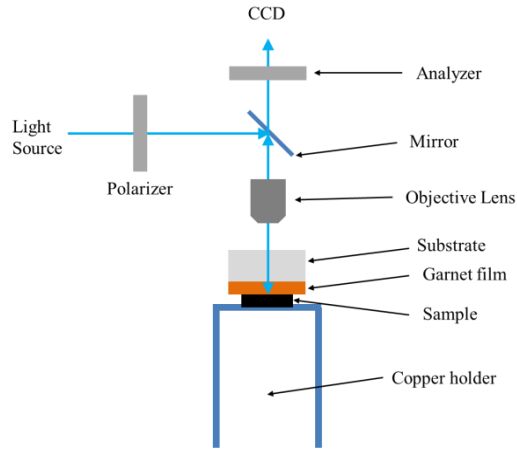

Figure S3. A schematic figure of the MO instrument.

## 1.4 AC magnetic susceptibility

A 3 mm × 3.5 mm × 400 nm (thickness of the superconducting layer) tape was placed and clamped at the 12.9 cm mark using two shorter PTFE tubes within a longer PTFE tube. Both shorter PTFE tubes and the tape were positioned inside the longer PTFE tube. The longer PTFE tube was then connected to the PPMS sample rod and placed inside the sample chamber for AC susceptibility measurements.

First, the AC magnetic susceptibility was measured while varying the DC field strength with an AC field amplitude of 2 Oe at 107 Hz. Next, the frequency of the AC field was varied while maintaining an AC field amplitude of 2 Oe and a DC field of 1 T. Finally, the AC field amplitude was changed while the DC field was set to 0 T and the AC field frequency was maintained at 107 Hz.

This allowed us to obtain the real part of the magnetic moment $m'$ and the imaginary part of the magnetic moment $m''$ of the tape under different conditions.

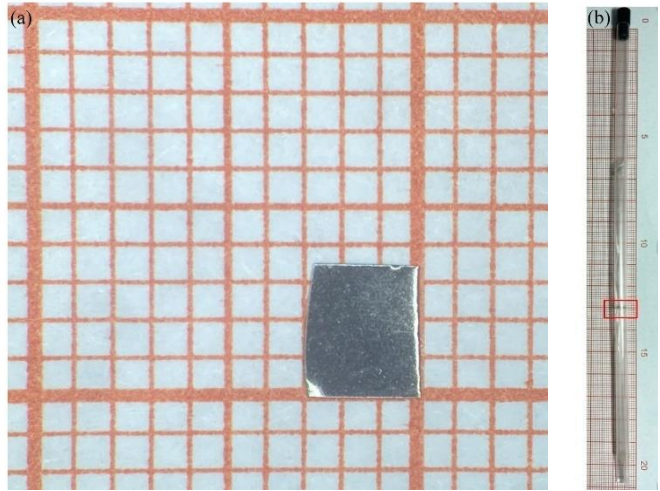

Figure S4. (a) FeTe$_{0.5}$Se$_{0.5}$ superconducting tape. (b) Polytetrafluoroethylene (PTFE) tube, with the tape located at 12.9 cm (within the red box).

## 2. Calculation of AC Loss
### 2.1 AC magnetic susceptibility

The real part of the magnetic moment $m'$ and the imaginary part of the magnetic moment $m''$ of the sample can be obtained through the ACMS component of the PPMS. The sample's AC susceptibility can then be calculated using the formulas:

$$\chi' = m' / (V \times H_{AC})$$

and

$$\chi'' = m'' / (V \times H_{AC})$$

where $V$ is the sample volume and $H_{AC}$ is the amplitude of the AC field.

### 2.2 AC Loss

According to the Bean model, magnetization $M$ is a function of the external field and is not directly related to time. In a periodic external field $H$, $M$ also varies periodically with time. The energy loss per unit volume during one cycle is represented by the area of the hysteresis loop, leading to the following formula:

$$Q = \oint \mu_0 H dM = \pi \mu_0 H_{AC}^2 \chi''.$$

When considering the energy loss per unit time, i.e., the AC hysteresis loss power, it is necessary to multiply by the frequency $f$. Thus, the following equation can be obtained:

$$P = \pi \mu_0 H_{AC}^2 \chi'' f$$

where $P$ is the AC hysteresis loss power.

# 3. Experimental Results
## 3.1 *I-V*

A set of *I-V* curves was obtained through transport measurements, as shown in Figures S5(a)-(l). The temperature range was from 3 K to 14 K, with data measured under magnetic fields of 0-9 T for each temperature. In Figure S5(m), the offset method is employed. The critical current is determined by the current at which a tangential line from part of the current-voltage curve crosses zero voltage.

As shown in Figure S5(n), the I-V curve was fitted using $V \propto I^n$. At $T$ = 11 K and 9 T, the superconducting transition region aligns well with $V \propto I^{16}$. Similarly, the inset displays *n* values from 3 K to 14 K under 9 T, with *n* reaching 52 at 4 K.

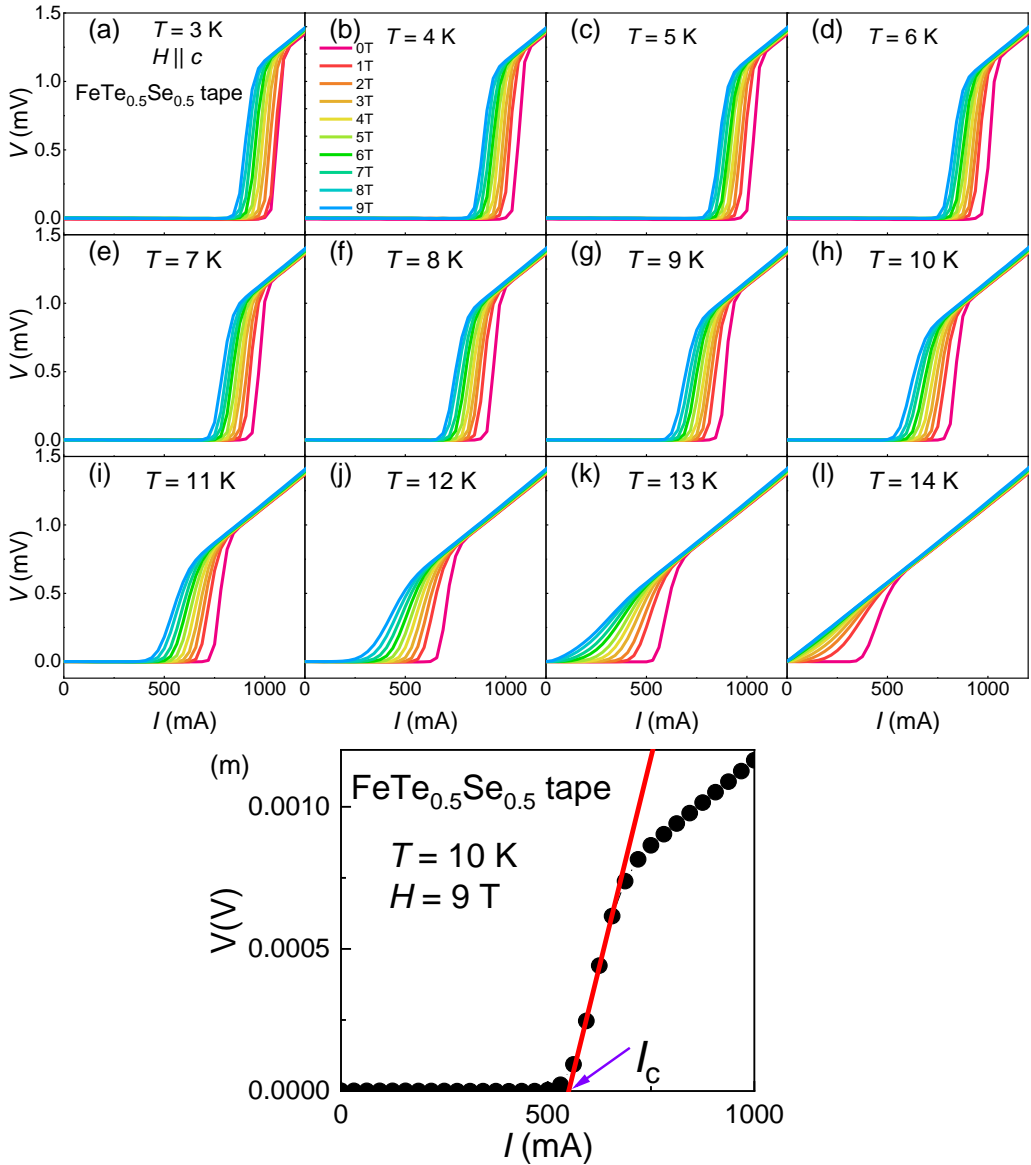

Figure S5. Voltage - current curves of FeTe$_{0.5}$Se$_{0.5}$ tape under different magnetic fields ($H \parallel c$) at (a) 3 K, (b) 4 K, (c) 5 K, (d) 6 K, (e) 7 K, (f) 8 K, (g) 9 K, (h) 10 K, (i) 11 K, (j) 12 K, (k) 13 K and (l) 14 K. (m) Off-set method.

## 3.1 *R-T*

Figures S6(a) and S6(b) display the resistance versus temperature curves for $H \parallel c$ and $H \parallel ab$, respectively. The first derivative of resistance with respect to temperature is shown in Figures S6(c) and S6(d). In this study, the magnetic field at the peak position is taken as the upper critical field at that temperature, while the magnetic field at the low-temperature endpoint of the peak is taken as the irreversible field at that temperature.

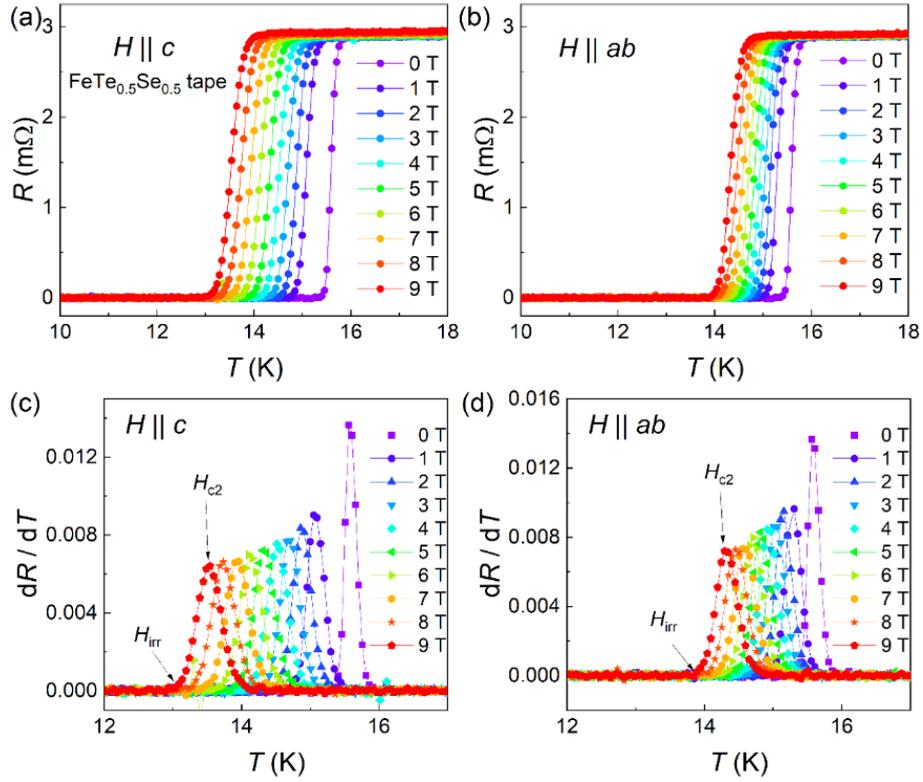

Figure S6. Temperature dependence of resistance under different magnetic fields for (a) $H \parallel c$ and (b) $H \parallel ab$ for the FeTe$_{0.5}$Se$_{0.5}$ tape. Under (c) $H \parallel c$ (d) $H \parallel ab$, the first derivative of resistance with respect to temperature.

### 3.3 MO

Figures S7 show the MO images of the FeTe$_{0.5}$Se$_{0.5}$ tape at temperatures of 11 K, 13 K, and 15 K, respectively. The formation of a double-Y shaped structure with an angle of 45º is clearly observed. As the temperature approaches $T_c$, no pronounced defects appear, confirming the good homogeneity of this tape.

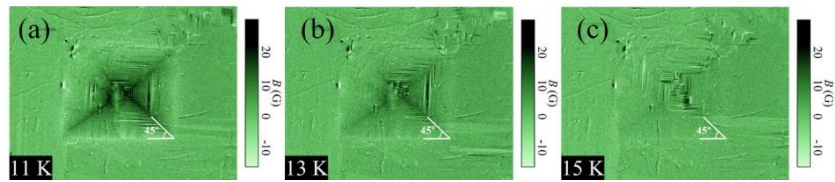

Figure S7. MO images of FeTe$_{0.5}$Se$_{0.5}$ tape in the remanent state at (a) $T = 11$ K, (b) $T = 13$ K, and (c) $T = 15$ K.

### 3.4 ACMS

According to the formula $Q = \oint \mu_0 H dM = \pi\mu_0 H_{AC}^2 \chi''$, it is evident that at the peak of $\chi''$, $\chi_p''$ corresponds to the peak value $Q_p$. We calculated $Q_p$ at different $H_{AC}$ values and different frequencies, as shown in Figures S8(a) and S8(b). As illustrated in Figure S8(a), due to the minimal variation in $\chi_p''$, $Q_p$ exhibits an approximate quadratic relationship with $H_{AC}$, which is consistent with the Bean model. Additionally, when applying AC fields at different frequencies, the $Q_p$ value for the tape remains nearly unchanged over one cycle, as shown in Figure S8(b), which is also reasonable if the flux creep in the sample is negligible.

Taking the frequency $f$ into account, the peak value of the AC hysteresis loss power, $P_p$, corresponding to $Q_p$, was obtained using the formula $P = \pi\mu_0 H_{AC}^2 \chi'' f$. Clearly, $P_p$ shows a distinct linear proportionality with $f$, as shown in Figure S9(b).

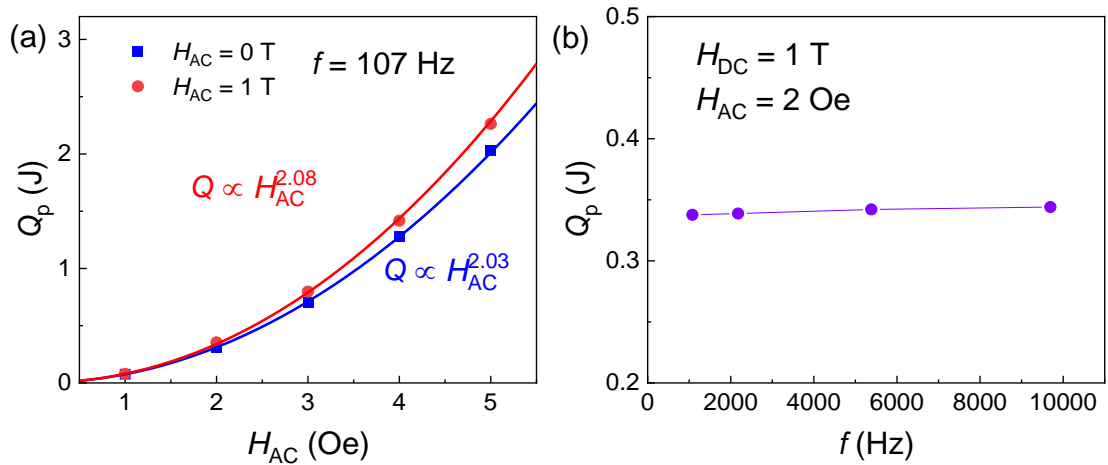

Figure S8. (a) The Relationship between $Q_p$ and $H_{AC}$. (b) The Relationship between $Q_p$ and $f$.

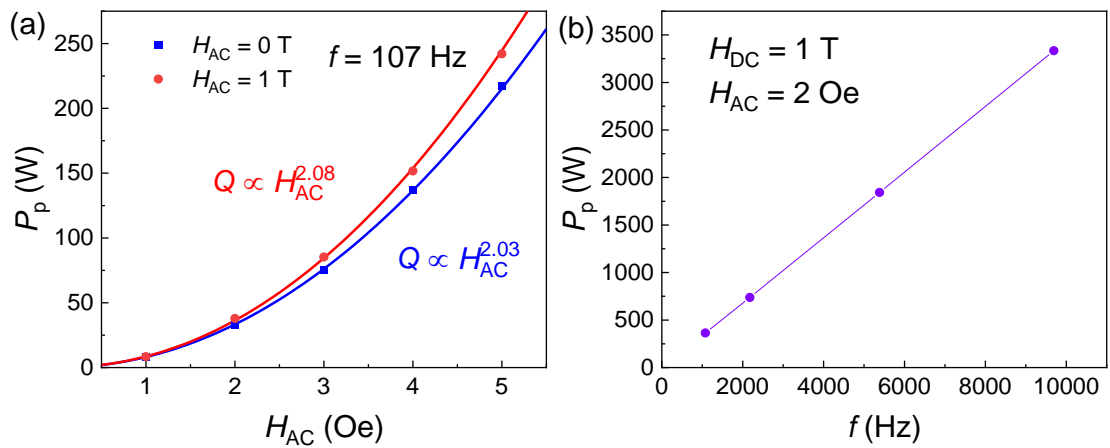

Figure S9. The Relationship between $P_p$ and $H_{AC}$. (b) The Relationship between $P_p$ and $f$.